\documentstyle[aps,prl,twocolumn]{revtex}

\draft

\begin{document}

\title{ Elementary
 excitations in trapped Bose 
 gases beyond mean field approximation}
\author{L. Pitaevskii$^{1,2,3,4}$ and S. Stringari$^{3,4}$}
\address{Department of Physics, Technion$^1$, 32000 Haifa, Israel}
\address{Kapitza Institute for Physical Problems$^2$,
117454, Moscow, Russia}
 \address{Dipartimento  di Fisica, Universit\`{a} di Trento$^3$,}
\address{and Istituto Nazionale di Fisica della Materia$^4$,
I-3850 Povo, Italy}
 \date{ July 23 1998}

 \maketitle

 \begin{abstract}
Using hydrodynamic theory of superfluids and the Lee-Huang-Yang
equation of state for interacting Bose gases we derive the
first correction to  the collective frequencies
of a trapped gas, due to 
 effects beyond mean field approximation. The corresponding frequency
shift, which is calculated at zero temperature and for large $N$,
 is compared with other corrections due
to finite size, non-linearity and finite temperature.
We show that for reasonable choices of the relevant parameters of 
the system, the non-mean field correction is the leading
contribution and amounts to about $1$\%. The role of the deformation
of the trap is also discussed. 

 \end{abstract}


\narrowtext

The experimental measurements \cite{bec,mit2,mit3} 
of the collective oscillations
of Bose condensed gases confined in magnetic traps have 
provided an excellent confirmation of the predictions 
of mean field theory (see \cite{rmp} for a recent theoretical review).
The accuracy of the mean field predictions is not surprising since
in these gases the average distance between particles is
significantly larger than the range of interatomic forces.
Typically, the gas parameter $n(0)a^3$, where $n(0)$ 
is the density evaluated
in the center of the trap, and $a$ is the $s$-wave scattering length,
is smaller than $10^{-4}$. According to the theory of Lee, Huang and 
Yang (LHY) \cite{LHY}, the first corrections to the mean field predictions
are expected to behave like $\sqrt{a^3n}$ and to be consequently of the order
of $1$\% or less in these systems. While such corrections are too
small to be observed in the density profiles or in the release
energy, they might be observable in the frequency of
the collective excitations where the accuracy of measurements is much 
higher. For example an accuracy of $0.3\sim 0.4$\% has been already achieved
in the experiment of \cite{mit3}. 

Measuring effects beyond mean field theory is a challenging task
and would open new perspectives in the many-body investigation
of these novel systems. So far the theoretical investigation of these 
effects has been limited
to the equilibrium properties,
either including first quantum corrections 
in analytic form
\cite{tommasini,nieto}, or
through
numerical 
simulations based on Monte Carlo methods \cite{krauth}. 
The purpose of this work is to
provide
an analytic calculation of the first corrections to the 
collective frequencies due to non-mean field effects. These
corrections are calculated  in the large
$N$ limit and at zero temperature. 

Our starting point are the hydrodynamic equations of  superfluids (see
for example \cite{NP})
 \begin{equation}
 \frac{\partial}{\partial t} n + \mbox{\boldmath$\nabla$}
({\bf v}n) = 0
 \label{continuity}
 \end{equation}
and
\begin{equation}
m \frac{\partial}{\partial t} {\bf v} +
\mbox{\boldmath$\nabla$}( \mu + \frac{1}{2}m{\bf
v}^2) = 0
\label{Euler}
 \end{equation}
holding at zero temperature. Here $n$ is the density of the system,
 ${\bf v}$ is the velocity field and $\mu$ is the chemical 
potential. Equations (\ref{continuity}-\ref{Euler}) 
permit to describe  the low frequency  collective excitations 
also in non-uniform Bose superfluids, provided 
 the density
profile varies on a macroscopic scale and one can  use the local density
approximation for the chemical potential
 \begin{equation}
 \mu({\bf r},t) = \mu_l(n({\bf r},t)) + V_{ext}({\bf r}) 
 \label{lda}
 \end{equation}
where $\mu_l(n)$ is the chemical potential calculated for a 
uniform gas at density
$n$ and $V_{ext}$ is the external confining potential.

In the following we will consider the linearized  regime of 
eqs.(\ref{continuity})
and (\ref{Euler}). We write  $n({\bf r},t)=n({\bf r}) + \delta n({\bf r},t)$ 
and $\mu({\bf r},t)
= \mu_0 + \delta \mu_l({\bf r},t)$ 
with $\delta\mu_l=(\partial\mu_l/\partial n)\delta n$, so that
eqs.(\ref{continuity})
and (\ref{Euler}) can  be rewritten in the useful form 
\begin{equation}
m{\partial^2 \delta  n \over \partial t^2} - 
\mbox{\boldmath$\nabla$}\left(n\mbox{\boldmath$\nabla$}
(\frac{\partial \mu_l}{\partial n}\delta n)\right) = 0 \, .
\label{HD}
\end{equation}
The ground state density $n({\bf r})$ 
entering equation (\ref{HD}) can be easily
calculated by imposing the equilibrium condition 
$\mu_0 = \mu_{\ell}(n({\bf r})) + V_{ext}({\bf r})$ 
where $\mu_o$ is the ground state value of the chemical potential, 
fixed to ensure the proper normalization of $n({\bf r})$.

Equations (\ref{continuity}-\ref{HD}) 
do not necessarily require that the trapped  gas is 
weakly interacting.
It is also worth noticing that the density $n$ entering these equations
 should not be confused with the
density  of the condensate, which in general does not obey
equations of macroscopic type. 
Only in the weakly interacting
limit, where quantum depletion effects are negligible, can the density of
the system
be identified with the condensate density. In this case  
eqs. (\ref{continuity}-\ref{HD}) are equivalent to  
the time dependent Gross-Pitaevskii
equations for the order parameter, 
evaluated in the large $N$ limit (see, for example, \cite{ss}).

According to LHY theory, the
chemical potential of a uniform interacting Bose gas is determined 
by the low density expansion:
 \begin{equation}
 \mu_l(n) = gn\left(1 + \frac{32}{3\sqrt\pi}\sqrt{a^3n}\right)
 \label{mul}
 \end{equation}
where $g=4\pi\hbar^2a/m$ is the interaction coupling constant
and $a$ is the $s$-wave scattering length. 
Eq.(\ref{mul}) represents a major result 
of many body theory and accounts
for non trivial renormalization effects of the interaction coupling constant.
It provides the first correction to the result $\mu=gn$ given by  lowest
order theory,  
hereafter called Bogoliubov or mean field approximation.
The LHY equation of state (\ref{mul}) can be derived
starting from Bogoliubov theory. In this scheme the energy of the system, 
including the zero-point motion of elementary excitations,  
is given by
$E = \frac{1}{2}Ngn + \frac{1}{2}\sum_{{\bf p}\ne 0}[\epsilon(p)
-p^2/2m -gn ]$
where $\epsilon(p)$ is the 
energy of elementary excitations in Bogoliubov theory.  The 
zero point energy contains 
an ultraviolet divergency at large $p$ which is cured by the 
proper renormalization 
$g \to g(1+g\frac{1}{V}\sum_{{\bf p}\ne 0}m/p^2)$ 
of the coupling constant, so that 
one finally obtains a convergent result for the energy of the system,
yielding result (\ref{mul}) for the chemical potential.
It is worth noticing
that the LHY correction 
does not involve additional parameters with respect to Bogoliubov theory, 
being 
fixed by the scattering length  and by the density. 

Using the LHY equation of state (\ref{mul}), the equation 
for the
ground state density 
can  be solved by iteration and one finds
the result \cite{tommasini}
\begin{equation}
n({\bf r}) = n_{TF}({\bf r}) - \alpha n_{TF}^{3/2}({\bf r})
\label{n0}
\end{equation}
where 
$n_{TF}({\bf r})=(\mu_0-V_{ext}({\bf r}))/g$ 
is the so called  Thomas-Fermi result for the ground 
state density \cite{baym} and
$\alpha = (32/3\sqrt\pi)a^{3/2}$. 

Notice that in the same scheme the condensate density $n_c$
is given by the 
expression 
$n_c({\bf r}) = n_{TF}({\bf r}) - 
\frac{5}{4}\alpha n_{TF}^{3/2}({\bf r})$
which is  smaller than the density $n({\bf r})$
and yields \cite{rmp} 
the result 
\begin{equation}
\frac{N_{out}}{N}=
\frac{5\sqrt{\pi}}{8}\sqrt{a^3n(0)}
\label{Nout}
\end{equation}
for the  
quantum depletion of the condensate of an atomic gas confined
in a harmonic trap. 

Using (\ref{mul}) one can write
$(\partial\mu_l/\partial n) = g(1 +3/2\alpha n_{TF}^{1/2})$ so that 
eq.(\ref{HD}) takes the  form
\begin{equation}
m \omega^2\delta n + 
\mbox{\boldmath$\nabla$}\left(gn_{TF} \mbox{\boldmath$\nabla$}
\delta n\right) = - \frac{1}{2}\mbox{\boldmath$\nabla$}^2\left(\alpha g  
n_{TF}^{3/2}\delta n\right) .
\label{HD1}
\end{equation}
Equation (\ref{HD1}) provides the appropriate generalization
of the zero-th order hydrodynamic equation 
\begin{equation}
m \omega^2\delta n + 
\mbox{\boldmath$\nabla$}\left(gn_{TF} \mbox{\boldmath$\nabla$}
\delta n\right) = 0
\label{HDS}
\end{equation}
used in \cite{ss} to evaluate the collective frequencies in the 
large $N$, Thomas-Fermi
approximation. Eq.(\ref{HDS}) admits analytical solutions if the external 
potential is harmonic. In particular, for an  isotropic 
trap ($V_{ext}=\frac{1}{2}m\omega_o^2r^2$) the solutions of (\ref{HDS})
obey the dispersion relation 
\cite{ss}
\begin{equation}
\omega(n_r,\ell) = \omega_0(2n_r^2 + 2n_r\ell + 3n_r + \ell)^{1/2}.
\label{dispersionS}
\end{equation} 
where $n_r$ is the number of  radial nodes and $\ell$ is the angular
momentum of the excitation.  
Equation (\ref{dispersionS}) shows that the collective frequencies
in the mean field approximation are fixed, apart from geometrical
factors, only by the oscillator frequency, a quantity measured with 
very high precision in experiments. This is a remarkable feature 
exhibited by these harmonically trapped gases which allows for a 
safe investigations of small corrections.

Once the solutions of (\ref{HDS}) are known, 
eq.(\ref{HD1}) can be easily solved 
by treating its right hand side as a small perturbation.
One finds that the corresponding  frequency shifts  
obey the general equation 
\begin{equation}
\frac{\delta \omega}{\omega} = -\frac{\alpha g}{4m\omega^2}
\frac{\int d^3{\bf r} (\mbox{\boldmath$\nabla$}^2 \delta n^*)\delta n 
n_{TF}^{3/2}}
{\int d^3{\bf r}  \delta n^*\delta n}
\label{deltaomega}
\end{equation}
where $\delta n$ are the solutions of (\ref{HDS}) and $\omega$ 
are the corresponding frequencies. The integrals of (\ref{deltaomega})
extend to the region where the Thomas-Fermi density is positive.

 In the absence of trapping the gas is uniform and the 
solutions of eqs. (\ref{HD1}-\ref{HDS}) have the form
$\delta n \sim e^{iqz}$ and exhibit 
a phonon dispersion
$\omega=cq$. In this case Eq.(\ref{HD1}) 
(or, equivalently, (\ref{deltaomega})) gives
the Beliaev result\cite{beliaev}
$\delta c/c = 8\sqrt{a^3n/\pi}$
for the shift of the sound velocity with respect to the
Bogoliubov value $c=\sqrt{gn/m}$, calculated
at the density $n$.  Notice that even in the 
uniform case $n$ differs from $n_{TF}$ because
of eq.(\ref{n0}). The shift of the sound velocity is 
consistent with the change in the compressibility 
$mc^2=n\partial\mu/\partial n$ associated with the  LHY
correction in the equation of state 
(\ref{mul}).

Eq.(\ref{deltaomega}) shows that the so called  "surface" oscillations 
$\delta n = r^{\ell}Y_{\ell m}$, satisfying the condition
 $\mbox{\boldmath$\nabla$}^2 \delta n=0$,
are not affected by the LHY correction. Physically this behavior
follows from the fact that in the long wavelength limit the hydrodynamic
surface oscillations are entirely determined by the external field.
For spherical trapping these  solutions 
obey the dispersion law
$\omega = \sqrt\ell \omega_0$ which is simply obtained setting $n_r=0$
in (\ref{dispersionS}).

In order to observe effects beyond
mean field one has consequently to focus on compressional modes, which
are  sensitive to the equation of state. The lowest mode 
in a spherical trap is the monopole
(breathing) oscillation ($n_r=1, \ell=0$), characterized by 
the zero-th order dispersion 
$\omega = \sqrt5 \omega_0$ and by density oscillations of the form 
$\delta n \sim (r^2-3/5R^2)$. 
 In this case Eq.(\ref{deltaomega}) 
yields the relevant result
\begin{equation}
\frac{\delta\omega_M}{\omega_M}= 
\frac{63\sqrt\pi}{128}\sqrt{a^3n(0)}
\label{monop}
\end{equation}
showing that the fractional  
shift of the monopole frequency is  proportional
to the square root of  the gas 
parameter evaluated  in the center of the trap. 
This correction exhibits the same dependence on the gas 
parameter as the quantum depletion of the condensate, 
although the coefficient of proportionality slightly
differs in the two cases. 
It is useful to write
the gas parameter
 in terms of the relevant parameters
of the system as \cite{rmp}
\begin{equation}
a^3n(0) = 
\frac{15^{2/5}} 
{8\pi}\left(N^{1/6}
\frac{a}{a_{ho}}\right)^{12/5}
\label{gasparameter}
\end{equation}
where $N$ 
is the number of atoms in the trap and 
$a_{ho}=(\hbar/m\omega_0)^{1/2}$ is the oscillator length.
Using, for example, $N=10^6$ and $a/a_{ho}=6\,10^{-3}$, we predict a
relative shift of $1$ \%. 
A similar value is found for the quantum depletion of the condensate.
Eq.(\ref{gasparameter}) shows that in order to enhance the value of the gas
parameter it is more effective to increase the value of the ratio $a/a_{ho}$
rather than the value of $N$ which enter the equation with a much 
lower power. In practice, however, it is not easy to obtain
large values of  $a^3n(0)$ and hence large
 frequency shifts. So far the achievement of high
densities  is in fact limited by  three body 
recombinations which cause the atoms to leave the trap.

The above shift of the monopole  frequency should be compared with other 
corrections which might be relevant in actual experiments, like finite
size, non-linearity and thermal effects. Finite size effects arise because even
in the mean field scheme the Thomas-Fermi value $\sqrt5\omega_0$ holds only in
the large $N$ limit. These corrections arise from the "quantum pressure term"
in the equation for the velocity field  which is ignored
in Eq.(\ref{Euler}), and 
 can be calculated
by a proper perturbation procedure in the Gross-Pitaevskii equation.
Using a
sum rule approach one obtains the following  result for 
the leading correction to the monopole
oscillation in the large $N$
limit and isotropic trapping \cite{zambelli}
\begin{equation}
\frac{\delta\omega_M}{\omega_M}= 
-\frac{7}{6}
\left(\frac{a_{ho}}{R}\right)^4\log\left(\frac{R}{Ca_{ho}}\right)
\label{monopsurf}
\end{equation}
where $C=1.3$ is a dimensionless parameter and 
$R  =  a_{ho}(15 N a/a_{ho})^{1/5}$  is
the radius of the system. 
For the surface quadrupole mode one finds that 
the fractional shift has opposite sign and
is larger by a factor $5$.

It is worth noticing that the corrections to the Thomas-Fermi value due 
to non-mean (\ref{monop}) and finite size (\ref{monopsurf})
effects 
depend on  different combinations of the relevant parameters of the system.
In fact finite size effects depend on the combination $Na/a_{ho}$, while non-mean
field effects depend on $N^{1/6}a/a_{ho}$. In the thermodynamic 
limit,  where $N\to \infty$ and $\omega_0 \to 0$, with the product
$N\omega^3_0$ kept constant,  finite size corrections go to zero, while
the gas parameter (\ref{gasparameter}) has a finite  value (for a discussion
of this limit in trapped Bose gases see, for example, \cite{rmp}).
The large $N$, thermodynamic limit is reasonably well 
realized in experiments. For example,
using the same values for $N$ and $a/a_{ho}$ employed above, 
one finds that the finite size shift (\ref{monopsurf}) of the monopole 
frequency
 is much smaller 
($\sim 0.1 $ percent) than the non-mean field correction (\ref{monop}). 
In general finite size effects are negligible if 
the condition $N \gg (a_{ho}/a)^2\log(R/a_{ho})$ is satisfied.

Non-linearity is another important effect to discuss. In 
fact in actual experiments the amplitude of the oscillation
cannot be made arbitrarily small. The 
effects of non-linearity have been investigated in details in 
\cite{minniti} in the framework of
the Thomas-Fermi approximation. The leading corrections to the 
frequency shift can be written in the
 form
$\delta\omega/\omega=  A^2 \delta$
where $A$ is the fractional amplitude of the oscillation 
of the atomic cloud
 confined 
in the trap and the coefficient $\delta$ can be calculated
in an explicit way \cite{minniti}. For the monopole mode in the 
spherical trap one has $\delta = 
- 1/6 $ so that for fractional  amplitudes  less than $10$ \%, the 
effects of non-linearity are very small.

In addition to finite size and non linearity effects, one should
also take into account that experiments are carried out at finite temperature.
 At present
there is no fully reliable theory to account for the temperature
dependence of the
collective frequencies of
these trapped gases. However one expects that these effects should 
vanish very rapidly when $kT$ is smaller than the chemical potential.
A rough estimate of the thermal effect can be obtained by assuming that 
the shift of the real part of $\omega$ is of the same order as 
its imaginary part which is 
responsible for the damping of the oscillation. 
This might provide an experimental 
control of the thermal effect on the frequency
shift also in the absence of accurate low temperature thermometry.

Finally an important question  concerns the role of the anisotropy
of the confining potential. In fact most of magnetic trap are at
present non spherical. Furthemore, high density values are at present more 
easily reached working with  highly deformed traps.
For an axially deformed trap 
of the form
$V_{ext} = {1\over2}m\omega_{\perp}^2r_{\perp}^2+{1\over 2}m\omega_z^2z^2$,
where $r_{\perp}=(x^2+y^2)^{1/2}$ is the radial
coordinate, 
the  HD equations (\ref{HDS}) admit several interesting solutions.
In addition to the dipole (center of mass) oscillation, the  excitations 
so far investigated experimentally
are the $m=2$ quadrupole  mode, whose density varies as $\delta n \sim
r^2Y_{22}$, and the $m=0$ oscillations resulting from the 
coupling between the quadrupole and the
monopole modes (notice that 
the $z$-th component, $m$, of angular momentum 
is still a good quantum number in axially deformed systems). The $m=2$
quadrupole mode has frequency $\omega = 2\omega_{\perp}$
and is not afffected by non-mean field effects since  
$\mbox{\boldmath$\nabla$}^2 \delta n =0$. Conversely the decoupled
frequencies of the $m=0$ modes are given by \cite{ss}
\begin{equation}
s^2 = \frac{\omega^2}{\omega^2_{\perp}}= 2 +\frac{3}{2}\lambda^2 \pm
\frac{1}{2} \left(9\lambda^4 - 16 \lambda^2 + 16 \right)^{1/2}
\label{s2}
\end{equation}
where $\lambda = \omega_z/\omega_{\perp}$ characterizes the asymmetry of the
trap. The corresponding density oscillations  have the form
$\delta n \sim \left(-2\mu (s^2-2)/m\omega^2 + r^2_{\perp} +
(s^2-4)z^2\right)$.
After some length, but straightforward algebra, one derives the 
result
\begin{equation}
\frac{\delta\omega}{\omega}= 
\frac{63\sqrt\pi}{128}\sqrt{a^3n(0)}f_{\pm}(\lambda)
\label{deltaomegadef}
\end{equation}
for the frequency shift of the $m=0$ modes where 
$$f_{\pm}(\lambda) = \frac{5}{3} 
\frac{(s^2-2)^2}{(3s^4-20s^2+40)} = \frac{1}{2} \pm 
\frac{8+\lambda^2}{6\sqrt{9\lambda^4 - 16 \lambda^2 + 16 }}$$
and  the index $\pm$ refers to the higher ($+$) and lower ($-$) 
solutions
of Eq.(\ref{s2}).
Notice that for a spherical trap ($\lambda=1$) 
the solutions of (\ref{s2}) are 
$s^2=5$ (monopole) and $s^2=2$ 
(quadrupole). In the 
first case one finds $f_+=1$ and one 
recovers result (\ref{monop}) for the breathing mode, while 
in the second case there is no shift. 
In the figure we show the   functions $f_+$ and $f_-$ relative to the two modes
as a function 
of the deformation parameter $\lambda$. 

Another important consequence of the use of deformed trap concerns
the effects of non-linearity. In fact it has been shown \cite{minniti}
that these effects can be amplified or reduced by changing the 
value of $\lambda$. For example, choosing $\lambda = 1.40$ 
one finds that the non-linearity coefficient $\delta$
of the high lying solution of (\ref{s2})  vanishes. 
For this mode 
one finds  $s^2 = 7.1$ and  $f_+ = 0.88$.
Another interesting case is obtained choosing $\lambda=\sqrt8$.
In this case the non-linear effect is vanishingly small for
the low energy solution.  
For this mode one finds $s^2=3.2$ and
$f_-=0.38$.

As already pointed out the largest 
values of the gas parameter have been 
so far reached
for
cigar trap configurations.
 In this case the frequency shift of
the lower  mode with dispersion $s^2=5\lambda^2/2$ is reduced 
significantly ($f_-(0)=1/6$) with respect to the case of the
$\lambda=1$ breathing mode ($f_+(1)=1$). Conversely the higher mode,
  which corresponds to a compressional radial oscillation with dispersion  
$s^2 =4$, exhibits only a small reduction ($f_+(0)=5/6$).
It is also worth pointing out that for this mode  
the non-linear effect vanishes as  \cite{minniti}
$ \delta\omega/\omega = -0.0938 \lambda ^2A^2$. 
 Finite
size effects are 
also vanishing in the same limit. 
The fact that non linear and finite size effects 
vanish for $\lambda \to 0$ reflects the occurrence of a 
hidden symmetry \cite{hid} of the Gross-Pitaevskii equation 
characterizing the radial compressional mode in two-dimensional
gases and migth be used
to improve the accuracy
of experimental measurements.
One should however notice that,
for very small values of $\lambda$,
the trapped gases exhibit, in addition to  the radial excitations, 
also a low-lying branch  of axial modes  \cite{axial}. 
This could produce
 a "parametric" instability \cite{mechanics}
of the compressional radial oscillation   due
to decay into two or more axial excitations.
 
In conclusion in this letter we have 
derived the first corrections to the 
collective frequencies of trapped Bose gases arising from
effects beyond mean field theory. We have shown that with reasonable
choices of the parameters  these effects, although 
small, might be visible experimenatally. Their direct observation
would represent an important  achievement in the study
of many-body effects  associated with the occurrence of 
Bose-Einstein condenstaion.

We would like to thank E. Cornell, J. Dalibar, W. Ketterle
and D. Stamper-Kurn for very stimulating discussions.

Figure Caption. Functions 
$f_+$ and $f_-$ relative to the higher and lower
$m=0$ modes (see eqs.(\ref{s2}) and (\ref{deltaomegadef})), 
as a function of the deformation parameter $\lambda = 
\omega_z/\omega_{\perp}$.


\begin{references}


%\bibitem{bec}  M.H.~Anderson et al., Science, {\bf 269}, 198 (1995);
%K.B.~Davis et al. Phys. Rev. Lett. {\bf 75}, 3969 (1995);  C.C. Bradley
%et al., Phys. Rev. Lett. {\bf 78}, 985 (1997) (see also Phys. Rev. Lett.
%{\bf 75}, 1687 (1995).

\bibitem{bec} D.S. Jin et al., Phys. Rev. Lett. {\bf  77}, 420 (1996).

\bibitem{mit2} M.-O. Mewes et al., Phys. Rev. Lett. {\bf 77}, 988 (1996).

\bibitem{mit3} D.M. Stamper-Kurn et al., cond-mat/9805022.

\bibitem{rmp} F. Dalfovo et al., cond-mat/9806038.

\bibitem{LHY} T.D. Lee and C.N. Yang,   Phys. Rev.
{\bf 105}, 1119 (1957);
T.D. Lee, K. W
Huang, and C.N. Yang,  Phys. Rev.       
{\bf 106}, 1135 (1957).

\bibitem{tommasini}
E. Timmermans, P. Tommasini, and K. Huang,
Phys. Rev. A {\bf 55}, 3645
(1997).

\bibitem{nieto} E. Braaten, and A. Nieto,  Phys. Rev.           
B {\bf 56}, 14745 (1997).


\bibitem{krauth} W. Krauth Phys. Rev. Lett. {\bf 77}, 3695 (1996),
M. Holzmann, and W. Krauth, cond-mat/9807014.

\bibitem{NP} P.~Nozi\`eres and D.~Pines, {\it The Theory of Quantum
Liquids}
(Addison-Wesley, Reading, MA, 1990) Vol.II.

\bibitem{ss} S.~Stringari, Phys. Rev. Lett. {\bf 77}, 2360 (1996).

\bibitem{baym} G. Baym and C. Pethick, Phys. Rev. Lett.
{\bf 76}, 6 (1996).

\bibitem{beliaev} S.T. Beliaev, Soviet Physics JETP {\bf 34}, 299  
(1958). 

\bibitem{zambelli} F. Zambelli, and S. Stringari, to be published.

\bibitem{minniti} 
F. Dalfovo, C. Minniti, and L.P. Pitaevskii,
Phys. Rev. A {\bf 56}, 4855 (1997). 

\bibitem{hid} L.P. Pitaevskii.
Phys. Lett. A {\bf 221}, 14 (1996); 
L.P. Pitaevskii, and A. Rosch, Phys. Rev. A {\bf 55}, R853
(1997).

\bibitem{axial} M. Fliesser et al., Phys Rev. A {\bf 56}, R2533 (1997);
S. Stringari, Phys. Rev. A, in press.

\bibitem{mechanics}  L.D. Landau, and E.M. Lifshitz, {\it  Mechanics},
3-rd edition, Butterworth Heinemann, Oxford, 1987.


\end{references}
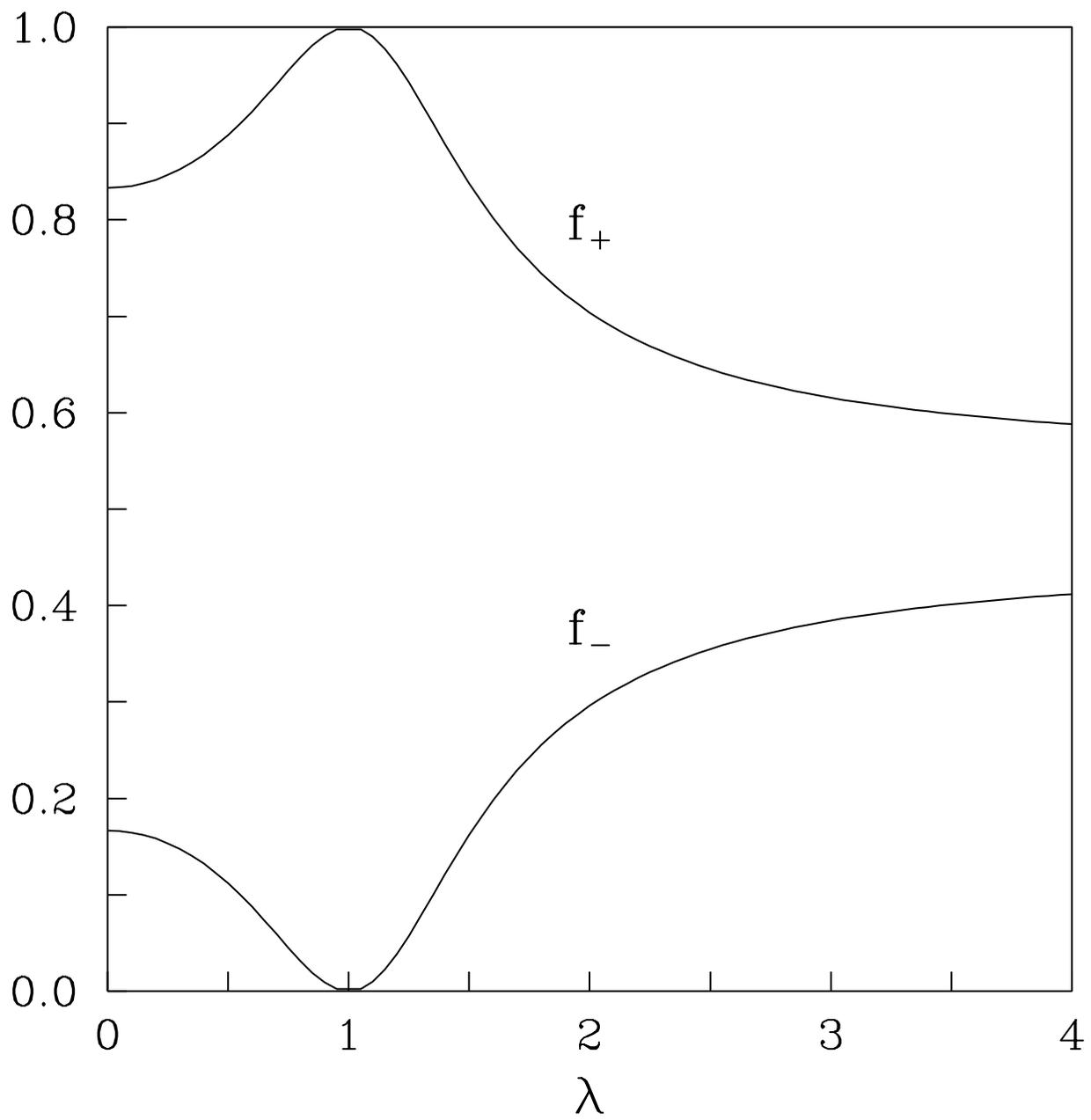
\end{document}